\documentclass[usenatbib]{basi}
\begin{document}
\title[Imaging polarimetry of the Bok globule CB56]{Imaging polarimetry of the Bok globule CB56}
\author[D.~Paul et al.]%
       {D. Paul$^1$, H.~S.~Das$^1$\thanks{email: \texttt{hsdas@iucaa.ernet.in}}
        and A. K. Sen$^1$\\
       $^1$Department of Physics, Assam university, Silchar 788 011, India\\
       }

\pubyear{2012}
\volume{40}
\status{in press}

\date{Received 2012 April 01; accepted 2012 June 06}

\maketitle
\label{firstpage}

\begin{abstract}
The measurement of polarization of the background stars in the region of Bok globules is 
important to study the magnetic field geometry and dust grain characteristics in the 
globule. These parameters are important for the formation and evolution of dark clouds. 
We made polarimetric observations of Bok globule CB56 in the R-filter from the 2-metre telescope 
at IUCAA Girawali Observatory (IGO). The observations were carried out on 2011 March 4th and 5th. 
The CCD images obtained from the instrument (IFOSC) were analyzed, to produce the polarization 
map of the Bok globule CB56.
\end{abstract}

\begin{keywords}
stars: formation -- ISM: clouds -- ISM: dust, extinction -- polarization
\end{keywords}

\section{Introduction}\label{s:intro}
Bok globules are the most simple molecular clouds in our Milky Way Galaxy, which are ideal sites for low-mass
star formation (Bok \& Reilly 1947). These are small, opaque, and relatively isolated molecular clouds with
diameters of about 0.7 pc (0.1 -- 2 pc) and masses of $\approx 10$ M$_\odot$ (2 -- 100 M$_\odot$) (Bok 1977,
and Leung 1985). Several catalogues of globules and dark clouds were published during the past years (e.g. 
Barnard 1927; Bok 1956; Lynds 1962; Sandquist \& Lindroos 1976; Feitzinger \& Stuewe 1984; Hartly et al. 1986; Clemens \& Barvainis 1988; Persi et al. 1990). The most homogeneous and complete compilation of 
dark clouds has been done by Clemens \& Barvainis (1988) (henceforth CB catalogue). This catalogue contains 248 small (mean size $\sim 4^\prime$) and nearby (distance $<$ 1 kpc) molecular clouds.

Magnetic fields play a major role on the evolution of dark clouds and may control the fragmentation of clouds to form stars (Mestel \& Spitzer 1956;
Nakano \& Nakamura 1978; Mouschovias \& Morton 1991; Li \& Nakamura 2004). Polarization arises from alignment
of interstellar dust grains in the presence of a magnetic field (Davis \& Greenstein 1951). When the light from
the background stars passes through the clouds, extinction and reddening are caused due to absorption and
scattering by the dust grains present in the clouds. This phenomenon also introduces linear polarization in the
starlight, if the dust grains are aligned and dichroic. Observations of the polarization of background stars
define the morphology of the magnetic field in the plane of the sky, and provide deep insights into the effect
of the field on the geometrical structure of a collapsing dark cloud (Goodman et al. 1990). Polarimetric
studies of these clouds provide important information about the optical properties of the dust grains as a function of environment.

There have been many studies of polarization with a view to trace the geometry of the magnetic fields in dark
clouds. Bhatt \& Jain (1993) presented linear polarization measurements of stars in the regions of the
molecular clouds B227 and L121. Bhatt \& Jain (1993) also discussed the magnetic field geometry in the clouds
as indicated by the polarization maps in relation to their morphology. Kane et al. (1995) studied Bok globule
CB4 using a CCD imaging polarimeter in order to create a detailed map of the magnetic field associated with CB4
cloud. Sen et al. (2000) have mapped eight star-forming clouds, CB3, CB25, CB39, CB52, CB54, CB58, CB62 and
CB246, in white light polarization and commented on the possible star formation dynamics there. Sen et al.
(2005) modelled the dark cloud as a simple dichroic polarizing sphere, which explains why polarization need not
always increase with total extinction A$_V$ as one moves towards the center of the cloud. Their analysis shows
that the observed polarization depends largely on the orientation of the magnetic field (within the cloud) with
respect to the direction of interstellar magnetic field. Ward-Thompson et al. (2009) studied optical and
submillimetre polarimetric data of the Bok globule CB3 and CB246. They found that the field orientation deduced
from the optical polarization data matches well with the orientation estimated from the submillimetre
measurements of the Bok globule CB3. Recently, Sen et al. (2010) studied three clouds CB3, CB25 and CB39
photopolarimetrically to find any possible relation between the observed polarization and extinction of the
stars in the background of these clouds. They found that the measured extinction values increase with the increase in percentage polarization for the cloud CB39 and to some extent for CB25. However, they did not observe any such correlation for cloud CB3.

In this paper, we present the results of polarization measurements for stars in the region of the Bok globule
CB56. This cloud is compact, irregular shaped   and has two infrared point sources in the IRAS Point Source
Catalogue (IRAS 07125-2503 and 07125-2507) associated with it (Clemens \& Barvainis 1988). CB56 is an ideal
site for star formation. Since no optical polarimetric observation of CB56 has yet been made, we select this
globule as our target object. It is to be noted that the polarimetric observations of dark cloud in the optical range help us to infer the magnetic field orientation in the low density edge regions of clouds, whereas the submillimetre polarimetric observation of dark cloud can only trace the field orientation in the high density central regions of the clouds.

\section{Observations}
\label{sect:obser}
Observations were made at the Cassegrain focus (f/10) of 2-m telescope of the Inter 
University Centre for Astronomy and Astrophysics (IUCAA) Girawali Observatory (IGO; 
lat.= 19{$^\circ$ 5$^\prime$N,  long.= + 73{$^\circ$ 40$^\prime$E, altitude = 1000m), Pune, 
India on 2011 March 4th and 5th. We report here optical polarization data of Bok globule CB56, from the catalogue of Clemens \& Barvainis (1988). The polarization was measured of stars background to the clouds and seen through the cloud periphery. Table 1 gives basic information of CB56 globule and Table 2 gives the observation log.

The focal plane instrument used was the IUCAA Faint Object Spectrograph and Camera (IFOSC). It employs an EEV
2K $\times$ 2K, thinned, back-illuminated CCD with 13.5$\mu m$ pixels. The CCD used for imaging provides an
effective field of view of 10.5$'$ $\times$ 10.5$'$ on the sky corresponding to a plate scale of 0.3 arcsec
pixel$^{-1}$. The gain and read out noise of the CCD camera are 1.5e$^{-}$/ADU and 4e$^{-}$ respectively.
IFOSC's capabilities are enhanced with imaging polarimetric mode with a reduced field of view of about 2
arcminute radius. It measures linear polarization in the wavelength range 350 -- 850 nm. It makes use of a
Wollaston prism and half-wave plate to observe two orthogonal polarization components that define a Stoke's
parameter. This removes the effects of sky polarization.   Two successive positions of the half-wave plate are
needed to retrieve the degree of polarization and the position angle of the polarization vector. The optical
principle and design of present IUCAA polarimeter are given by Scarrott et al. (1983) and Sen \& Tandon (1994). More details on the principles of the instrument are described in Ramaprakash et al. (1998). When a half-wave plate rotates successively 22.5$^\circ$, 45$^\circ$ and 67.5$^\circ$ from an initial position denoted 0, the position angles of the polarized components respectively rotate 45$^\circ$, 90$^\circ$ and 135$^\circ$ from the initial position. A CCD (2048 $\times$ 2048 pixels of 13.5$\mu m$ each) camera was used to image the data. The observations were carried out at R-filter with typically a 1200-sec exposure time for each object. For each object within the field of view, two images, called the ordinary and extraordinary, are formed on the CCD. These two images correspond to the two orthogonal linear polarization components that are measured simultaneously. The separation between the ordinary and extraordinary components is 0.8 arcmin. It is to be noted that our analysis is restricted to single wavelength, i.e., R-filter, because we are mainly interested in the average properties of the magnetic field pattern and not on the wavelength dependence of polarization.

For the purpose of calibration, we observed one polarized (HD 147084) and one unpolarized
standard star (HD 65583) taken from a list by Serkowski (1974). The detail observation
information  are presented in Table 3, where the observed polarization (P$_{obs}$) and position angle (PA$_{obs}$) are also shown. These values are close to the standard values. The data reduction of the FITS images which we get from IFOSC CCD, has been done by the IRAF (Image Reduction and Analysis Facility) software. A polarimetric package was developed for data analysis within the IRAF environments using a mixture of standard IRAF tasks and a FORTRAN programme.

\begin{table}
\caption{Basic information of CB56 globule.}
\begin{center}
\begin{tabular}{ c c c c c c }
\hline
   RA (2000) & DEC (2000)& Position & Galactic  & Galactic  & Dimension \\
             &           & angle ($\theta$)    & longitude ($l$) & latitude ($b$)& \\
 \hline
    07:14:36 & $-$25:08:54.2 & 170$^\circ$ & 237.9274 & -6.4587 & 4.5$^{\prime}$  $\times$ 2.2$^{\prime}$\\
 \hline
\end{tabular}
\end{center}
\end{table}
\begin{table}
\begin{center}
\caption{Observation log of CB56.}
\begin{tabular}{ c c c c c c }
\hline
   Date & UT & RA (2000)  & DEC (2000)   \\
 \hline
 2011 March 04 & 17:23 & 07:14:36 & $-$25:08:54.2\\
 2011 March 05 & 14:21 & 07:14:36 & $-$25:08:54.2\\
 \hline
\end{tabular}
\end{center}
\end{table}

\begin{table}
\caption{Observations of polarized and unpolarized standard stars on 2011 March 4 and 5.}
\begin{center}
\begin{tabular}{ c c c c c c c }
\hline
  Star          & RA (2000) & DEC (2000) & $P (\%)$ & PA ($^o$) & $P_{obs} (\%)$ & PA$_{obs}$ ($^o$)   \\
 \hline
 HD 147084     & 16:20:38.18 & -- 24:10:09.5 &4.46 & 32 & 4.18 &36\\
 \hline
 HD 65583      & 08:00:32.12 & + 29:12:44.4 & 0.013  & 144.7& 0.05& 149.3\\
 \hline
\end{tabular}
\end{center}
\end{table}
\begin{table}
\caption{Observed linear polarization values for various field stars in CB56.}
\begin{center}
\begin{tabular}{ c c c c c r r }
\hline
  Star  \#     & RA (2000) & DEC (2000) & P (\%) & E$_p$ (\%) & $\theta$ ($^\circ$)  & E$_\theta$ ($^\circ$)\\
 \hline
1 & 07:14:27.02 & $-$25:09:20.53&3.74&0.09& 157.5 & 0.7\\
2 & 07:14:30.64 & $-$25:07:19.68&0.95&0.17& 173.0 & 5.1\\
3 & 07:14:31.82 & $-$25:09:37.55&0.81&0.17& 157.5 & 6.0\\
4 & 07:14:33.36 & $-$25:11:01.10&0.98&0.21& 22.5  & 6.2\\
5 & 07:14:35.68 & $-$25:07:49.48&0.92&0.27& 0      & 8.4\\
6 & 07:14:38.00 & $-$25:06:38.51&0.65&0.10& 157.5 & 4.4\\
7 & 07:14:38.14 & $-$25:09:42.93&0.65&0.17& 157.5 & 7.5\\
8 & 07:14:38.98 & $-$25:10:17.06&0.52&0.19& 31.7  & 10.6\\
9 & 07:14:44.14 & $-$25:07:33.98&1.86&0.20& 149.9 & 3.1\\
 \hline
\end{tabular}
\end{center}
\end{table}

\begin{figure}
\centerline{\includegraphics[width=11cm]{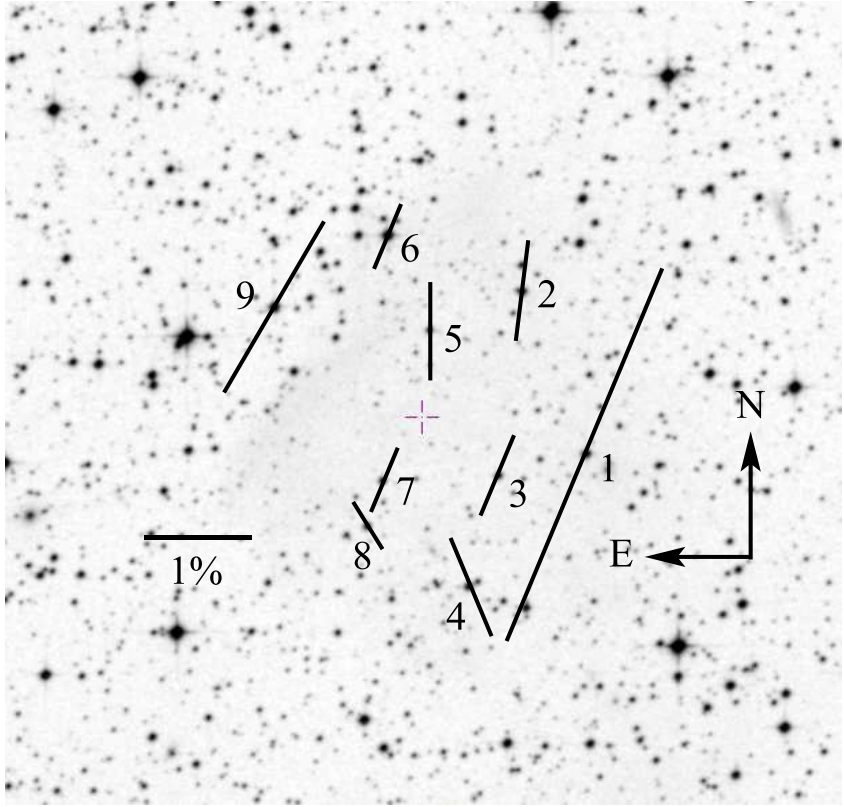}}
\caption{Polarization map of the cloud CB56 with a field of view 10.5$'$ $\times$ 10.5$'$. Star numbers are
listed in Table 4 (Field centre: RA = 07:14:36, DEC = $-$25:08:54.2). The vectors are the optical 
polarization vectors from Table 4. \label{f:one}}
\end{figure}

\section{Results}
In Table 4, the results of our polarization measurements for stars in the regions of CB56 are listed. The image
of CB56 with a field of view 10.5$'$ $\times$ 10.5$'$ has been taken from DSS and is shown in Fig.~1. The
observed stars have been numbered in order of increasing right ascension. For programme stars that could be
identified with stars in the DSS catalogue, we give the DSS numbers in column 1 of the Table 4. The other
columns give the polarization P (in \%), the position angle $\theta$ (in degrees) and the standard errors E$_p$
and E$_\theta$ associated with P and $\theta$. The position angle $\theta$ is measured from north increasing
eastward. In Fig.~1, centered on the stars observed, the polarization vectors have been drawn. The length of the polarization vector is proportional to the percentage polarization P and it is oriented in the direction indicated by $\theta$.

\section{Discussion}
The polarization map shown in Fig.~1 represents the geometry of the magnetic field, projected on to the plane of the sky, in the regions of the clouds observed.

CB56 is a compact, irregular shaped cloud (galactic co-ordinates: $l = 237.93 ^\circ$, $b = - 6.46^\circ$)
whose major axis as a position angle 170$^\circ$ and has two infrared point sources in the IRAS Point Source
Catalogue (IRAS 07125-2503 and 07125-2507) associated with it (Clemens \& Barvainis 1988). The polarization
vectors for most of the stars in Fig.~1 are nearly parallel to the long axis of the cloud. Position angles for only two stars (numbered 4 and 8) are out of a total of nine observed stars show polarization position angles that deviate significantly from the mean value. Thus excluding star numbers 4 and 8, the mean value for $<\theta>$ = 161.8$^\circ$ and the standard deviation $\sigma _\theta$ = 9.80$^\circ$. This value for $<\theta>$ is nearly the same as the position angle (170$^\circ$) of the long axis of the cloud. The mean value of the polarization $<P>$ for the 7 stars is 1.369\%, while the values of P for stars 4 and 8 are 0.977\% and 0.515\% respectively. These stars with lower values of P and $\theta$ very different from the mean value for the rest of the stars may be foreground to the cloud.

It would be interesting to see the submillimetre polarimetric measurements of the above globule which can confirm the magnetic field orientation in the cloud in addition to optical polarimetry. But no such measurement has been done yet on CB56 cloud.

\section{Conclusions}
We have presented the results of linear polarization measurement for stars in the region of the dark cloud CB56. The polarization map of this cloud indicates that the magnetic field in CB56 is more or less unidirectional and nearly parallel to the long axis of the cloud, and has the same direction as the local interstellar magnetic field (Clemens \& Barvainis 1988).

\section*{Acknowledgments}
The authors HSD and AKS acknowledge  Inter University Centre for
Astronomy and Astrophysics (IUCAA), Pune for its associateship
programme. The authors gratefully acknowledge the reviewers 
N. M. Ashok and Derek Ward-Thompson for their useful suggestions which 
definitely helped to improve the quality of the paper.


\label{lastpage}
\end{document}